\documentclass[PRC,aps,showpacs,twocolumn,superscriptaddress,floatfix]{revtex4-1}

\usepackage{ulem}

\usepackage[dvipdfmx]{graphicx}
\usepackage{here}
\usepackage{amsmath,amssymb,times}
\usepackage{color}

\newcommand{\beq}{\begin{equation}}
\newcommand{\eeq}{\end{equation}}
\newcommand{\bea}{\begin{eqnarray}}
\newcommand{\eea}{\end{eqnarray}}

\renewcommand{\L}{\Lambda}

\def\gsim{~\,\makebox(1,1){$\stackrel{>}{\widetilde{}}$}\,~}
\def\lsim{~\,\makebox(1,1){$\stackrel{<}{\widetilde{}}$}\,~}

\begin{document}

\title{
Chiral $g$-matrix folding-model  approach to reaction cross sections \\
for scattering of Ca isotopes on a C target}

\author{Shingo~Tagami}
\affiliation{Department of Physics, Kyushu University, Fukuoka 812-8581, Japan}

\author{Masaomi~Tanaka}
\affiliation{Department of Physics, Kyushu University, Fukuoka 812-8581, Japan}

\author{Maya~Takechi}
\affiliation{Niigata University, Niigata 950-2181, Japan}

\author{Mitsunori~Fukuda}
\affiliation{Department of Physics, Osaka University, Osaka 560-0043, Japan}

\author{Masanobu~Yahiro}
\email[]{orion093g@gmail.com}
\affiliation{Department of Physics, Kyushu University, Fukuoka 812-8581, Japan}

\begin{abstract}
We first predict the ground-state properties of Ca isotopes, using the Gogny-D1S Hartree-Fock-Bogoliubov (GHFB) with and without the angular momentum projection (AMP).   
We find that $^{64}$Ca is an even-dripline nucleus and $^{59}$Ca is an odd-dripline nucleus, using 
$A$ dependence of the one-neutron separation energy  $S_{1}$ and 
the two-neutron separation energy, $S_{2}$. 
As for $S_{1}$, $S_{2}$ and the binding energies $E_{\rm B}$, our results agree with the experimental data  
in $^{40-58}$Ca.    
As other ground-state properties of $^{40-60,62,64}$Ca, we predict 
charge, proton, neutron, matter radii, neutron skin and deformation. 
As for charge radii, our results are consistent with the experimental data in $^{40-52}$Ca. 
For $^{48}$Ca, our results on proton, neutron, matter radii agree with the experimental data. 
Very lately, Tanaka~{\it et. al.} measured 
interaction cross sections for $^{42-51}$Ca  scattering on  a $^{12}$C target 
at an incident energy per nucleon of $E_{\rm lab}=280$~MeV. 
Secondly, we predict  reaction cross sections $\sigma_{\rm R}$ for $^{40-60,62,64}$Ca, 
using a chiral  $g$-matrix double-folding model (DFM).  
To show the reliability of the present DFM for $\sigma_{\rm R}$, we apply  the DFM for the data
on $^{12}$C scattering on  $^{9}$Be, $^{12}$C, $^{27}$Al targets 
in  $30  \lsim E_{\rm lab} \lsim 400 $~MeV, and show that the present DFM is good 
in $30  \lsim E_{\rm lab} \lsim 100 $~MeV and $250  \lsim E_{\rm lab} \lsim 400 $~MeV.  
For $110  \lsim E_{\rm lab} \lsim 240 $~MeV, our results have small errors. 
To improve the present DFM for $\sigma_{\rm R}$, we propose two prescriptions.  
\end{abstract}

\maketitle

\section{Introduction}
\label{Sec:Introduction}

Systematic understanding of unstable nuclei is a goal in nuclear physics. 
In fact, neutron-rich nuclei near the neutron-drip line are synthesized in nature 
by the r process. In particular, the binding energies $E_{\rm B}$ affect the synthesis; see the homepage 
NuDat 2.7~\cite{HP:NuDat 2.7} 
for the measured values. 
The odd and the even dripline are determined from mass-number ($A$) 
dependence of 
the one-neutron separation energy  $S_{1}(A) \equiv E_{\rm B}(A)-E_{\rm B}(A-1)$ and 
the two-neutron separation energy 
$S_{2}(A) \equiv E_{\rm B}(A)-E_{\rm B}(A-2)$; see  Refs.~\cite{HP:NuDat 2.7,Tarasov2018,Michimasa2018obr,Neufcourt:2019qvd} 
for the experimental data. 

In many papers using the Glauber model, nuclear matter radii $r_{\rm m}$ are 
extracted from interaction cross sections $\sigma_{\rm I}$ and reaction cross sections 
$\sigma_{\rm R}$ ($\sigma_{\rm R} \approx \sigma_{\rm I}$); see 
Refs.~\cite{Tanihata:1986kh, AlKhalili:1996gm, Horiuchi:2006ga, Takechi:2009zz, Takechi:2012zz,Takechi:2014eza, Kanungo:2016tmz,Togano:2016wyx} as important  papers.  
Particularly for halo nuclei, the $r_{\rm m}$ are determined for $^6$He, $^8$B, $^{11}$Li, $^{11}$Be 
in Refs.~\cite{Tanihata:1986kh, AlKhalili:1996gm}, $^{19}$C in  Ref.~\cite{Kanungo:2016tmz}, 
$^{22}$C in Refs.~\cite{Horiuchi:2006ga,Togano:2016wyx} and $^{37}$Mg in Ref.~\cite{Takechi:2014eza}.  
We proposed a parameter quantifying the halo nature 
of one-neutron nuclei~\cite{Yahiro:2016edm}; see Fig. 3 of Ref.~\cite{Yahiro:2016edm} for 
seeing how halo the nucleus is.

High precision measurements of $\sigma_{\rm R}$ within 2\% error  were made 
for  $^{12}$C scattering on $^{9}$Be, $^{12}$C, $^{27}$Al targets in a wide range of incident energies~\cite{Takechi:2009zz}; 
say $30  \lsim E_{\rm lab} \lsim 400 $~MeV for  
$E_{\rm lab}$ being the incident energy per nucleon. In fact,  
the $r_{\rm m}$ of  $^{9}$Be, $^{12}$C, $^{27}$Al were 
determined by the Glauber model.  
Very lately,  in RIKEN,  Tanaka~{\it el. al.} measured  $\sigma_{\rm I}$ for $^{42-51}$Ca  
scattering on  a $^{12}$C target at $E_{\rm lab}=280$~MeV~\cite{Tanaka:2019pdo}.

The reliability of the Glauber model was investigated by constructing 
the multiple scattering theory
for nucleus-nucleus scattering~\cite{Yahiro:2008dr}. 
The eikonal approximation used in the Glauber model is not good for nucleon-nucleon collision 
in nucleus-nucleus scattering; see Fig. 1 of Ref.~\cite{Yahiro:2008dr}.
This problem can be solved by formulating the Glauber model with the multiple scattering theory. 
The formulation shows that  the nucleon-nucleon collision should be described by the $g$ matrix 
for lower energies and by the $t$ matrix for higher energies. 
The Glauber model is thus justified for higher $E_{\rm lab}$, say in $E_{\rm lab} \gsim 150$~MeV.

The $g$-matrix DFM~\cite{Brieva-Rook,Amos,CEG07,Minomo:2011bb, Sumi:2012fr, Egashira:2014zda,Watanabe:2014zea,Toyokawa:2017pdd} is a standard way of deriving microscopic optical 
potentials of nucleus-nucleus elastic scattering. The $g$-matrix DFM is thus a standard method 
for calculating $\sigma_{\rm R}$.   
The microscopic potentials are obtained by folding the $g$ matrix with projectile and target densities.   
In fact, the potentials have been used for elastic scattering in many papers. 
Using the DFM with  the Melbourne $g$-matrix, we discovered that $^{31}$Ne is a halo nucleus 
with strong deformation~\cite{Minomo:2011bb},   
and determined, with high accuracy, the  $r_{\rm m}$ for Ne isotopes~\cite{Sumi:2012fr} and  for Mg isotopes~\cite{Watanabe:2014zea},

As for the symmetric nuclear matter, Kohno calculated the $g$ matrix 
by using the Brueckner-Hartree-Fock (BHF) method with chiral N$^{3}$LO 2NFs and NNLO 3NFs~\cite{Koh13}. 
The BHF energy per nucleon becomes minimum at $\rho = 0.8\rho_{0}$ for the 
cutoff scale $\Lambda = 550$~MeV~\cite{Koh:Err}, 
 if the relation $c_D \simeq 4c_E$ is satisfied,  where $\rho$ is nuclear matter density and 
 $\rho_{0}$ stands for  the normal density.  
He then took $c_D=-2.5$ and $c_E=0.25$ so that  the energy per nucleon may be minimum 
at $\rho = \rho_{0}$.   Eventually, a better saturation  curve was obtained. 
The framework is applied for positive energies. The resulting non-local chiral  $g$ matrix is localized 
into three-range Gaussian forms by using the localization method proposed 
by the Melbourne group~\cite{von-Geramb-1991,Amos-1994,Amos}. 
We refer to the resulting local  $g$ matrix as Kyushu  
$g$-matrix in this paper~\cite{Toyokawa:2017pdd}.

As an {\it ab initio} method for structure of Ca isotopes,
we can consider the coupled-cluster 
method~\cite{Hagen:2013nca,Hagen:2015yea} 
with chiral interaction. 
Chiral interactions have been constructed by 
two groups~\cite{Weinberg:1991um, Epelbaum:2008ga, Machleidt:2011zz}.   
Among the effective interactions, NNLO$_{\rm sat}$~\cite{Ekstrom:2015rta} is 
the next-to-next-to-leading order chiral interaction that is constrained 
by radii and binding energies of selected nuclei up to $A \approx 25$~\cite{Hagen:2015yea}. 
In fact, 
the {\it ab initio} calculations were done 
for Ca isotopes~\cite{Ekstrom:2015rta, Hagen:2015yea, Ruiz:2016gne}.  
Garcia Ruiz {\it et. al.} evaluated the charge radii $r_{\rm ch}$ for 
$^{40-54}$Ca~\cite{Ruiz:2016gne}, 
using the coupled-cluster method with two low-momentum effective interactions, 
SRG1~\cite{Hebeler:2010xb} and SRG2~\cite{Furnstahl:2013oba}, 
derived from the chiral  interaction with the renormalization group method.

Particularly for a neutron-rich double-magic nucleus  $^{48}$Ca, the neutron skin 
$r_{\rm skin}=r_{\rm n}-r_{\rm p}$ was 
directly determined from a high-resolution measurement of $E1$ polarizability in RCNP~\cite{Birkhan:2016qkr}, where 
$r_{\rm n}$ and $r_{\rm p}$ are the rms radii of neutron and proton distributions, respectively.  
The value $r_{\rm skin}=0.14$--0.20~fm is important to determine not only 
the equation of state but also 
$r_{\rm n}$ and $r_{\rm m}$ of $^{48}$Ca. 
Using $r_{\rm p}=$3.40~fm~\cite{C12-density,Angeli:2013epw,Ong:2010gf}  evaluated from 
the electron scattering, 
we can find that $r_{\rm n}=3.54$--3.60=3.57(3)~fm and $r_{\rm m}=3.48$--3.52=3.50(2)~fm.

In this paper, we predict the ground-state properties of Ca isotopes 
using the Gogny-D1S Hartree-Fock-Bogoliubov (GHFB) 
with and without the angular momentum projection (AMP)~\cite{Tagami-AMP}, 
and predict $\sigma_{\rm R}$ for scattering of Ca isotopes  on a $^{12}$C target 
at  $E_{\rm lab}=280$~MeV by taking the Kyushu $g$-matrix DFM~\cite{Toyokawa:2017pdd}. 
The GHFB with and without the AMP are referred to as  ``GHFB+AMP'' and ``GHFB'', respectively. 
Details of our predictions are shown below.  

As an essential property of Ca isotopes, we first determine 
the odd and even driplines for Ca isotopes by seeing $A$ dependence of 
$S_{1}$ and $S_{2}$, and find that  $^{64}$Ca is an even-dripline nucleus 
and $^{59}$Ca is an odd-dripline nucleus. 
As for $E_{\rm B}$, $S_{1}$, $S_{2}$, our results 
agree with the experimental data 
for $^{40-58}$Ca~\cite{HP:NuDat 2.7,Tarasov2018,Michimasa2018obr,Neufcourt:2019qvd}. 
Our results are thus accurate enough for the prediction on the odd and even driplines.  

As other grand-state properties, we consider  
$r_{\rm ch}$, $r_{\rm p}$, $r_{\rm n}$, $r_{\rm m}$, 
$r_{\rm skin}$, deformation for $^{40-60,62,64}$Ca. 
As for the charge radii $r_{\rm ch}$, our results 
are consistent with the data~\cite{Angeli:2013epw} determined from 
the isotope shift method based on the electron scattering 
in $^{40-52}$Ca. 
As for $r_{\rm p}$, $r_{\rm n}$, $r_{\rm m}$,  $r_{\rm skin}$, the experimental data 
are available for   $^{48}$Ca~\cite{Birkhan:2016qkr}, and our results agree with  the data. 
The success on the ground-state properties indicates that the densities calculated with 
GHFB and GHFB+AMP are reliable for Ca isotopes.

The  Kyushu $g$-matrix folding model is successful in reproducing the differential cross sections 
of $p$ scattering at $E_{\rm lab}=65$~MeV~\cite{Toyokawa:2014yma} and of $^4$He scattering at 
$E_{\rm lab}=30 \sim 200$~MeV~\cite{Toyokawa:2015zxa,Toyokawa:2017pdd}. 
However, it is not clear whether the Kyushu $g$-matrix DFM is reliable for $\sigma_{\rm R}$. 
In order to investigate the reliability, we apply the Kyushu $g$-matrix DFM for measured $\sigma_{\rm R}$ 
on $^{12}$C scattering 
from $^{9}$Be, $^{12}$C, $^{27}$Al targets 
in $30  \lsim E_{\rm lab} \lsim 400 $~MeV, and 
confirm that the present DFM is reliable 
in $30  \lsim E_{\rm lab} \lsim 100 $~MeV and $250  \lsim E_{\rm lab} \lsim 400 $~MeV. 
We then predict $\sigma_{\rm R}$ for scattering of Ca isotopes  on a $^{12}$C target 
at  $E_{\rm lab}=280$~MeV, using the Kyushu $g$-matrix DFM~\cite{Toyokawa:2017pdd}. 
The reason for this prediction is that (1) 
the data on $\sigma_{\rm R}$ for $^{42-51}$Ca will be available soon and (2) 
the densities are determined accurately for Ca isotopes.

The present DFM is not accurate enough for $^{12}$C-$^{12}$C scattering 
in  $110  \lsim E_{\rm lab} \lsim 240$~MeV. 
In order to improve the present DFM in $110  \lsim E_{\rm lab} \lsim 240$~MeV, 
we propose two prescriptions. 

We explain our framework in Sec.~\ref{Sec:Theoretical framework}.
Our results are shown in Sec.~\ref{Sec:Results}.
Section~\ref{Sec:summary} is devoted to a summary.

\section{Framework}
\label{Sec:Theoretical framework}

Our framework is composed of GHFB and GFHB+AMP for structure and 
the Kyushu $g$-matrix DFM for reaction.

We determine the ground-state properties of Ca isotopes, using GHFB and GHFB+AMP~\cite{Tagami-AMP}. 
In GHFB+AMP, the total wave function  $| \Psi^I_{M} \rangle$ with the AMP is defined by 
\begin{equation}
 | \Psi^I_{M} \rangle =
 \sum_{K, n=1}^{N+1} g^I_{K n} \hat P^I_{MK}|\Phi_n \rangle ,
\label{eq:prjc}
\end{equation}
where $\hat P^I_{MK}$ is the angular-momentum-projector and the 
$|\Phi_n \rangle$ for $n=1,2,\cdots,N+1$ are mean-field (GHFB) states, 
where $N$ is the number  of the  states that one can block.  
The coefficients $g^I_{K n}$ are determined
by solving the following Hill-Wheeler equation,
\begin{equation}
 \sum_{K^\prime n^\prime }{\cal H}^I_{Kn,K^\prime n^\prime }\ g^I_{K^\prime n^\prime } =
 E_I\,
 \sum_{K^\prime n^\prime }{\cal N}^I_{Kn,K^\prime n^\prime }\ g^I_{K^\prime n^\prime },
\end{equation}
with the Hamiltonian and norm kernels defined by
\begin{equation}
 \left\{ \begin{array}{c}
   {\cal H}^I_{Kn,K^\prime n^\prime } \\ {\cal N}^I_{Kn,K^\prime n^\prime } \end{array}
 \right\} = \langle \Phi_n |
 \left\{ \begin{array}{c}
   \hat{H} \\ 1 \end{array}
 \right\} \hat{P}_{KK^\prime }^I | \Phi_{n'} \rangle.
\end{equation}

For odd nuclei,  we have to put a quasi-particle in a level, but the number $N$ of the blocking states 
are quite large. It is not easy to solve  the Hill-Wheeler equation with large $N$. 
Furthermore, we have to confirm that the resulting  $| \Psi^I_{M} \rangle$ converges with respect to 
increasing $N$ for any set of two deformations  $\beta$ and $\gamma$. This procedure is quite time-consuming. For this reason, we do not consider the AMP for  odd nuclei. 
As for GHFB, we consider the one-quasiparticle state that yields the lowest energy, so that 
we do not have to solve the Hill-Wheeler equation. However, it is not easy  to find the values of 
$\beta$ and $\gamma$ at which the energy becomes minimum in the $\beta$-$\gamma$ plane.

For even nuclei, there is no blocking state, i.e., $N=0$ in the Hill-Wheeler equation. 
We can thus consider GHFB+AMP. 
However,  we have to find the value of $\beta$ at 
which the ground-state energy becomes minimum.  
In this step, the AMP has to be performed for any $\beta$, so that the Hill-Wheeler calculation is still heavy. 
In fact, the AMP is not taken for most of mean field calculations; see for example Ref.~\cite{HP:AMEDEE}.  
The reason why we do not take into account $\gamma$ deformation is 
that the deformation does not affect  $\sigma_{\rm R}$~\cite{Sumi:2012fr}.

As a result of the heavy calculations for even nuclei, we find that $\beta$ is small for GHFB+AMP. 
Meanwhile, the mean-field (GHFB) calculations yield that the energy surface becomes minimum 
at $\beta=0$. The fact that $\beta=0$ for GHFB and small for GHFB+AMP 
yields small difference between GHFB results and GHFB+AMP ones; 
see Table~\ref{Deformation parameters for Ca isotopes.} 
for the values of $\beta$. In the table, we also show 
the values of $\beta$ and $\gamma$ for odd nuclei.

\begin{table}[htb]
\begin{center}
\caption
{Deformation parameters for Ca isotopes. 
 }
 \begin{tabular}{cccc}
  \hline
$A$ & $\beta^{\rm AMP}$ & $\beta$ & $\gamma$\\
  \hline
  40 &  0.093 & 0 &   \cr
  41 &  & 0.0320 & -180  \cr
  42 &  0.146 & 0 &  \cr
  43 &  &0.00976 & 60  \cr
  44 &  0.135 & 0 &  \cr
  45 &  & 0.0139 & 0.0599  \cr
  46 &  0.137 & 0 &  \cr
  47 &  & 0.00908 & -104  \cr
  48 &  -0.116 & 0 &  \cr
  49 &  & 0.0239 & 60  \cr
  50 & 0.121 & 0 &  \cr
  51 &  & 0.0199 & 8.94  \cr
  52 & -0.114 & 0 & \cr
  53 &  & 0.00173 & 0.0631 \cr
  54 & 0.130 & 0 &  \cr
  55 &  & 0.00195 & -177  \cr
  56 &  0.126 & 0 &  \cr
  57 &  & 0.000701 & -180  \cr
  58 &  -0.110 & 0 &  \cr
  59 &  & 0.0198 & 0.942  \cr
  60 &  0.111 & 0 &   \cr
  62 &  0.131 & 0 &   \cr
  64 &  0.138 & 0 &   \cr
  \hline
 \end{tabular}  
 \label{Deformation parameters for Ca isotopes.}
\end{center} 
\end{table}

We predict $\sigma_{\rm R}$ for scattering of $^{40-60,62,64}$Ca on a $^{12}$C target 
at  $E_{\rm lab}=280$~MeV, using the Kyushu $g$-matrix DFM~\cite{Toyokawa:2017pdd}. 
In the DFM, the potential $U$ between a projectile and a target is 
obtained by folding the  Kyushu $g$-matrix with the projectile and target densities;  
see Eq. (9) of Ref.~\cite{Toyokawa:2017pdd}. 
As for the densities, we adopt both GHFB and GHFB+AMP for even nuclei and 
GHFB for odd nuclei. As a way of making the center-of-mass correction, three methods were 
proposed in Refs. \cite{Tassie:1958zz,Horiuchi:2006ga,Sumi:2012fr}. We used 
the method of Ref.~~\cite{Sumi:2012fr}, since the procedure is quite simple.

As already mentioned in Sec. \ref{Sec:Introduction}, 
the present folding model is successful in reproducing the differential cross sections of $p$ scattering at 
$E_{\rm lab}=65$~MeV~\cite{Toyokawa:2014yma} and of $^4$He scattering at 
$E_{\rm lab}=30 \sim 200$~MeV~\cite{Toyokawa:2015zxa,Toyokawa:2017pdd}. 
To show the reliability of the present DFB for $\sigma_{\rm R}$, we apply  the present DFM for the data
on $^{12}$C scattering on  $^{9}$Be, $^{12}$C, $^{27}$Al targets 
in  $30  \lsim E_{\rm lab} \lsim 400 $~MeV, and show that the present DFM is good 
in $30  \lsim E_{\rm lab} \lsim 100 $~MeV and $250  \lsim E_{\rm lab} \lsim 400 $~MeV.  
For light nuclei $^{9}$Be, $^{12}$C, $^{27}$Al, we take the phenomenological densities~\cite{C12-density} 
deduced from the electron scattering; note that the phenomenological densities reproduce the experimental 
data~\cite{Takechi:2009zz} on $r_{\rm m}$.  
For the densities of even Ca isotopes, we take GHFB with $\beta = 0 $ and GHFB+AMP with $\beta$ deformation 
in order to investigate effects of $\beta$ deformation. 
As for the densities of odd Ca isotopes, we adopt GHFB in which  $\beta$ and $\gamma$ deformations are 
taken into account.

\section{Results}
\label{Sec:Results}

Using GHFB and GHFB+AMP, we first 
determine the odd (even) dripline of Ca isotopes by seeing the values of 
$S_{1}$ ($S_{2}$), and find  
that $^{64}$Ca is an even-dripline nucleus and $^{59}$Ca is an odd-dripline nucleus. 
For $^{40-60,62,64}$Ca, we then present the ground-state properties 
($E_{\rm B}$, $S_{1}$, $S_{2}$, $r_{\rm ch}$, $r_{\rm p}$, $r_{\rm n}$, $r_{\rm m}$, $r_{\rm skin}$, deformation). 
The theoretical results are consistent with the corresponding data. 
In the case that the experimental data are not available, 
we predict the the ground-state properties of   $^{40-60,62,64}$Ca, .

As stated in Sec. \ref{Sec:Introduction}, the Kyushu $g$-matrix folding model 
is successful in reproducing the differential cross sections 
of $p$ scattering at $E_{\rm lab}=65$~MeV~\cite{Toyokawa:2014yma} and of $^4$He scattering 
at $E_{\rm lab}=30 \sim 200$~MeV~\cite{Toyokawa:2015zxa,Toyokawa:2017pdd}. 
However, it is not clear whether the present DFM is reliable for $\sigma_{\rm R}$. 
We then apply the present DFM for measured $\sigma_{\rm R}$ on $^{12}$C scattering 
on $^{9}$Be, $^{12}$C, $^{27}$Al targets 
in $30  \lsim E_{\rm lab} \lsim 400 $~MeV, and 
show  that the present DFM is reliable 
in $30  \lsim E_{\rm lab} \lsim 100 $~MeV and $250  \lsim E_{\rm lab} \lsim 400 $~MeV. 
After confirming the reliability of the Kyushu $g$-matrix DFM, 
we predict $\sigma_{\rm R}$ for scattering of $^{40-60,62,64}$Ca on a $^{12}$C target at  $E_{\rm lab}=280$~MeV, since the data on $\sigma_{\rm R}$ will be available soon for $^{42-51}$Ca and the 
$r_{\rm m}$ are unknown for Ca isotopes except for $^{42,44,48}$Ca.  
The prediction is made with the GHFB densities, since we confirm 
that  effects of the AMP on $\sigma_{\rm R}$ are small.

\subsection{Determination of even and odd driplines for Ca isotopes}

We determine even and odd driplines, seeing $A$ dependence of  $S_{1}(A)$ and $S_{2}(A)$ and 
using the fact that nuclei are unbound for negative $S_{1}(A)$ and $S_{2}(A)$.

\begin{figure}[H]
\centering
\vspace{0cm}
\includegraphics[width=0.45\textwidth]{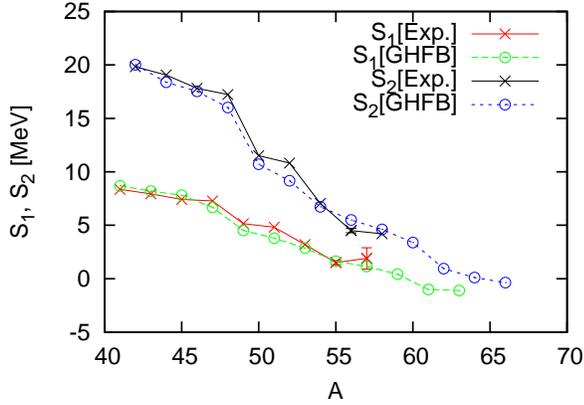}
\vspace{-10pt}
\caption{$A$ dependence of $S_{1}(A)$ and $S_{2}(A)$. 
Open circles show the GHFB  results for $S_{1}$ and $S_{2}$. 
Experimental data (crosses) are taken from Refs.~\cite{HP:NuDat 2.7,Tarasov2018,Michimasa2018obr}.  
}
\label{fig:$A$ dependence of S1S2}
\end{figure}

Figure~\ref{fig:$A$ dependence of S1S2} shows $S_{1}(A)$ and $S_{2}(A)$ as a function of $A$. 
The GHFB+AMP results are not plotted, since the results almost agree with the  GHFB results. 
The  GHFB results (open circles)  are consistent with the data (crosses)
on $S_{1}(A)$ and $S_{2}(A)$~\cite{HP:NuDat 2.7,Tarasov2018,Michimasa2018obr}. 
Seeing $A$ dependence of GHFB results, we can find that $^{64}$Ca is an even-dripline nucleus 
and $^{59}$Ca is an odd-dripline nucleus. 
The result is consistent with the observed line in Fig. 3 of Ref.~\cite{Neufcourt:2019qvd}.

\subsection{Binding energies of Ca isotopes}

\begin{figure}[H]
\centering
\vspace{0cm}
\includegraphics[width=0.45\textwidth]{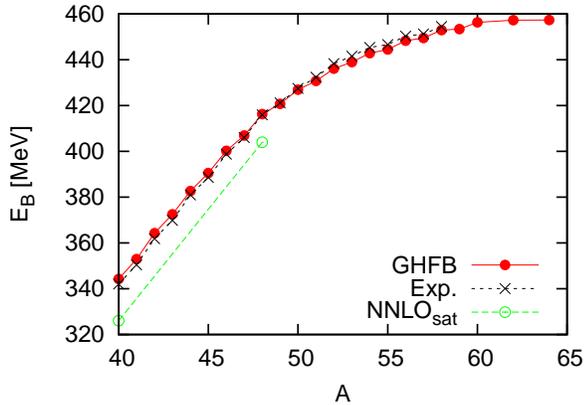}
\vspace{-10pt}
\caption{$A$ dependence of binding energy $E_{\rm B}(A)$. 
The GHFB results  are shown by closed circles. Open circles denote the results~\cite{Hagen:2015yea} of coupled-cluster calculations based on NNLO$_{\rm sat}$. 
Experimental data (crosses) are taken from 
the homepage NuDat 2.7~\cite{HP:NuDat 2.7}. 
}
\label{fig:$A$ dependence of BE}
\end{figure}

Figure \ref{fig:$A$ dependence of BE} shows $E_{\rm B}(A)$ as a function of $A$ from 40 to 64. 
The GHFB+AMP results are close to the GHFB ones (closed circles) for even Ca isotopes; 
in fact, the former deviates from the latter  at most by 0.73~\%. 
For this reason,  the GHFB+AMP results are not shown in Fig.~\ref{fig:$A$ dependence of BE}. 
The GHFB results reproduce the experimental data (crosses)
for $^{40-52}$Ca~\cite{HP:NuDat 2.7}, 
and yield better agreement with the experimental data than 
coupled-cluster results (open circles)~\cite{Hagen:2015yea} based on NNLO$_{\rm sat}$.

\subsection{Charge radii of Ca isotopes}
\label{Sec:Charge radii}

\begin{figure}[H]
\centering
\vspace{0cm}
\includegraphics[width=0.45\textwidth]{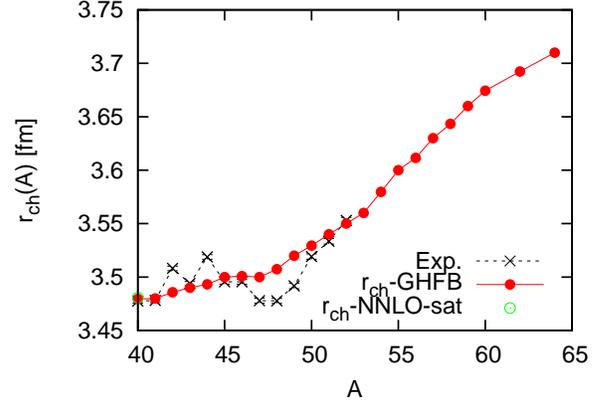}
\vspace{-10pt}
\caption{$A$ dependence of charge radii $r_{\rm ch}(A)$. 
The GHFB results  are shown by closed circles.  
Open circles denote the results~\cite{Hagen:2015yea} of coupled-cluster calculations based on NNLO$_{\rm sat}$ for $^{40}$Ca.
Experimental data (crosses) are taken from Ref.~\cite{Angeli:2013epw}. 
}
\label{fig:$A$ dependence of r-ch}
\end{figure}

Figure \ref{fig:$A$ dependence of r-ch} shows $r_{\rm ch}$ as a function of $A$.  
The GHFB+AMP results agree with the GHFB ones for even Ca isotopes; 
in fact, the former deviates from the latter  at most by 0.66~\%. 
For this reason,  the GHFB+AMP results are not shown in Fig.~\ref{fig:$A$ dependence of r-ch}. 
The GHFB results (closed circles) reproduce the experimental data (crosses)~\cite{Angeli:2013epw} 
derived from the isotope shift method based on the electron scattering for $^{40-52}$Ca; 
the former is deviated from the latter at most 0.9~\%.  
For $^{40}$Ca, the GHFB result agrees with 
 the result~\cite{Hagen:2015yea} (open circle) of coupled-cluster calculations based on NNLO$_{\rm sat}$.

\subsection{Radii and skin of Ca isotopes}
\label{Sec:Radii}

\begin{figure}[H]
\centering
\vspace{0cm}
\includegraphics[width=0.45\textwidth]{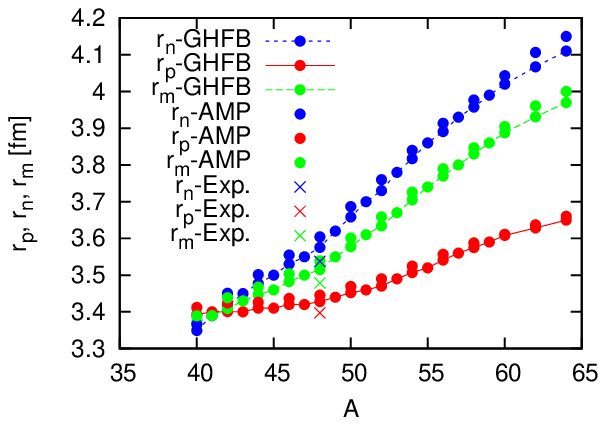}
\includegraphics[width=0.45\textwidth]{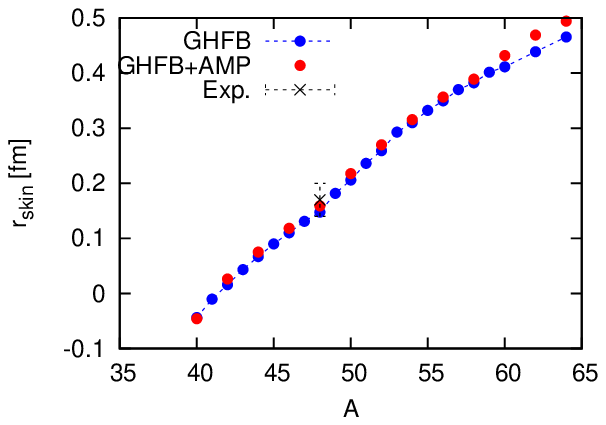}
\vspace{-10pt}
\caption{$A$ dependence of  $r_{\rm p}$, $r_{\rm n}$, $r_{\rm m}$ in the upper panel and 
$r_{\rm skin}$ in the lower panel. 
Tangent lines with closed  circles denote the GHFB results, while  closed circles correspond 
to the GHFB+AMP results.  
Experimental data (crosses) are taken from Ref.~\cite{Birkhan:2016qkr}. 
}
\label{fig:$A$ dependence of radii}
\end{figure}

Figure \ref{fig:$A$ dependence of radii} shows $r_{\rm p}$, $r_{\rm n}$, $r_{\rm m}$, $r_{\rm skin}$ 
as a function of $A$. 
The difference between GHFB+AMP (open circles) and GHFB (closed circles)  is small for even Ca isotopes;  
in fact, the former deviates from the latter  at most by 0.8~\% for $r_{\rm m}$. 
The reason for the small difference is that $\beta$ is small for  GHFB+AMP and zero for  GHFB, 
as shown in Table~\ref{Deformation parameters for Ca isotopes.}. 
Particularly for $^{48}$Ca, the experimental data are available~\cite{Birkhan:2016qkr}.  
The deviation of the GHFB+AMP result from the data (crosses) and  is 1.1 \% for $r_{\rm m}$. 
This indicates that the GHFB+AMP and GFHB are good enough for explaining the data. 
Our results on radii and skin are tabulated in Table~\ref{Radii for Ca isotopes. }. 

\begin{table}[htb]
\begin{center}
\caption
{Radii for Ca isotopes. 
The superscript ``AMP'' stands for the results of GHFB+AMP, and no
superscript corresponds to those of GHFB.
 }
 \begin{tabular}{ccccccccc}
  \hline
$A$ & $r_n^{\rm AMP}$ & $r_p^{\rm AMP}$ & $r_m^{\rm AMP}$ & $r_{\rm skin}^{\rm AMP}$ &
$r_n$ & $r_p$ & $r_m$ & $r_{\rm skin}$ \\
  \hline
  40 & 3.366 & 3.412 & 3.389 & -0.046 & 3.349 & 3.393 & 3.371  & \cr
  41 &  &  &  &  & 3.387 & 3.397 & 3.392 & -0.010  \cr
  42 & 3.451 & 3.424 & 3.438 & 0.026 & 3.417 & 3.401 & 3.409  & \cr
  43 &  &  &  &  & 3.448 & 3.405 & 3.428 & 0.043  \cr
  44 & 3.501 & 3.426 & 3.467  & 0.075 & 3.477 & 3.410 & 3.447 &  \cr
  45 &  &  &  &  & 3.504 & 3.414 & 3.465 & 0.090  \cr
  46 & 3.555 & 3.436 & 3.504 & 0.118 & 3.530 & 3.420 & 3.483 &  \cr
  47 &  &  &  &  & 3.554 & 3.424 & 3.499 & 0.131  \cr
  48 & 3.604 & 3.445 & 3.539 & 0.159 & 3.576 & 3.428 & 3.515 & \cr
  49 &  &  &  &  & 3.621 & 3.440 & 3.548 & 0.182  \cr
  50 & 3.687 & 3.469 & 3.601 & 0.218 & 3.658 & 3.452 & 3.577 &  \cr
  51 &  &  &  &  & 3.698 & 3.462 & 3.607 & 0.236 \cr
  52 & 3.760 & 3.490 & 3.659 & 0.270 & 3.734 & 3.475 & 3.577 &  \cr
  53 &  &  &  &  & 3.779 & 3.486 & 3.671 & 0.293 \cr
  54 & 3.840 & 3.524 & 3.726 & 0.316 & 3.817 & 3.507 & 3.705 &  \cr
  55 &  &  &  &  & 3.856 & 3.524 & 3.739 & 0.332  \cr
  56 & 3.913 & 3.557 & 3.790 & 0.357 & 3.891 & 3.541 & 3.770 &  \cr
  57 &  &  &  &  & 3.928 & 3.557 & 3.802 & 0.370  \cr
  58 & 3.977 & 3.588 & 3.847 & 0.389 & 3.958 & 3.575 & 3.830 &  \cr
  59 &  &  &  &  & 3.995 & 3.593 & 3.863 & 0.402  \cr
  60 & 4.043 & 3.611 & 3.904 & 0.432 & 4.020 & 3.608 & 3.888 &  \cr
  62 & 4.106 & 3.637 & 3.961 & 0.469 & 4.067 & 3.628 & 3.931 &  \cr
  64 & 4.153 & 3.658 & 4.005 & 0.494 & 4.113 & 3.648 & 3.974 &  \cr
  \hline
 \end{tabular}   
 \label{Radii for Ca isotopes. }
\end{center} \end{table}

\subsection{Prediction on  $\sigma_{\rm R}$ for $^{40-60,62,64}$Ca+$^{12}$C scattering at  $E_{\rm lab}=280$~MeV}
\label{Sec:Prediction}

\begin{figure}[H]
\centering
\vspace{0cm}
\includegraphics[width=0.45\textwidth]{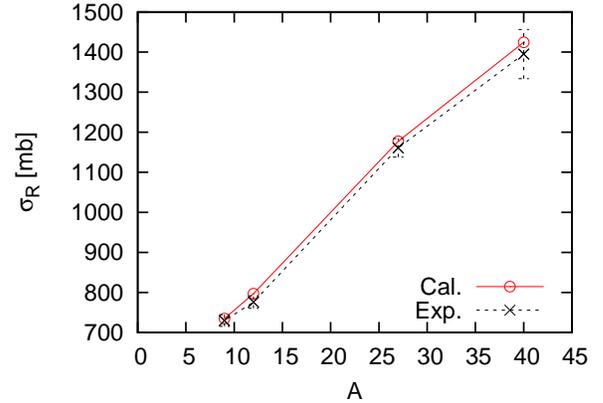}
\caption{Reaction cross sections $\sigma_{\rm R}$ for $^{12}$C scattering 
on  $^{9}$Be$, ^{12}$C, $^{27}$Al,  $^{40}$Ca targets at $E_{\rm n} =280$~MeV. 
Open circles denote the DFM results. 
The experimental data (crosses) are taken from Ref.~\cite{Takechi:2009zz}  
for $^{9}$Be$, ^{12}$C, $^{27}$Al  
and Ref.~\cite{Takechi:2014eza} for $^{40}$Ca; note that $E_{\rm lab}=250.7$~MeV for  $^{40}$Ca. 
}
\label{fig:C on Be-Ca-280}
\end{figure}

At first, we confirm the reliability of the present DFM for  $\sigma_{\rm R}$ at $E_{\rm lab} =280$~MeV, 
as seen in Fig.~\ref{fig:C on Be-Ca-280}. 
The DFM results (open circles) reproduce 
the experimental data (crosses)~\cite{Takechi:2009zz} for $^{9}$Be$, ^{12}$C, $^{27}$Al. 
Also for $^{40}$Ca, good agreement is seen between the DFM result with GHFB+AMP density (open circle) and 
the experimental data (cross); note that $E_{\rm lab}=250.7$~MeV for the  data~\cite{Takechi:2014eza}.

\begin{figure}[H]
\centering
\vspace{0cm}
\includegraphics[width=0.45\textwidth]{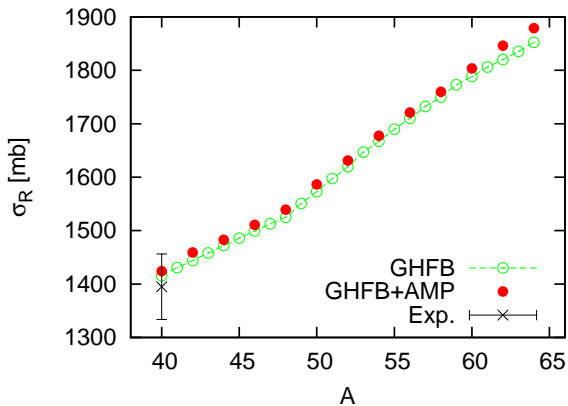}
\caption{
$A$ dependence of $\sigma_{\rm R}$ on $^{40-60,62,64}$Ca on a $^{12}$C target at  $E_{\rm lab}=280$~MeV. 
The DFM results with GHFB and GHFB+AMP densities are shown by open and closed circles, 
respectively. 
The experimental data~\cite{Takechi:2014eza} for $^{40}$Ca+$^{12}$C scattering at  $E_{\rm lab}=250.7$~MeV is 
denoted by a cross with error bar. 
}
\label{fig:$A$ dependence of RXSEC-Ca+C}
\end{figure}

Figure \ref{fig:$A$ dependence of RXSEC-Ca+C} is our prediction on $\sigma_{\rm R}$ 
for  $^{40-60,62,64}$Ca at  $E_{\rm lab}=280$~MeV. 
For $^{40}$Ca, the DFM results with GHFB and GHFB+AMP densities (open and closed circles) 
agree with the experimental data~\cite{Takechi:2014eza} at  $E_{\rm lab}=250.7$~MeV. 
The difference between the GHFB and GHFB+AMP densities is small. 
This comes from the fact that for even Ca isotopes the $\beta$ are zero for  GHFB and small for GHFB+AMP; see 
Table \ref{Deformation parameters for Ca isotopes.} for the values of $\beta$.

 . 
\subsection{Reaction cross sections in $30  \lsim E_{\rm lab} \lsim 400 $~MeV }

Through the analyses in Sec~\ref{Sec:Charge radii}$\sim$\ref{Sec:Prediction}, 
we can conclude  that the $\sigma_{\rm R}$ calculated with 
the Kyushu $g$-matrix DFM is valid for $^{40-60,62,64}$Ca+$^{12}$C scattering at  $E_{\rm lab}=280$~MeV. 
We then investigate how reliable the present DFM  is for  a wide range of $E_{\rm lab}$. 
For this purpose, we consider  
$^{12}$C scattering on  $^{9}$Be$, ^{12}$C, $^{27}$Al targets in 
$30 \lsim E_{\rm lab} \lsim 400$~MeV, since 
high-quality data are available~\cite{Takechi:2009zz}.

\begin{figure}[H] 
\centering
\vspace{0cm}
\includegraphics[width=0.45\textwidth]{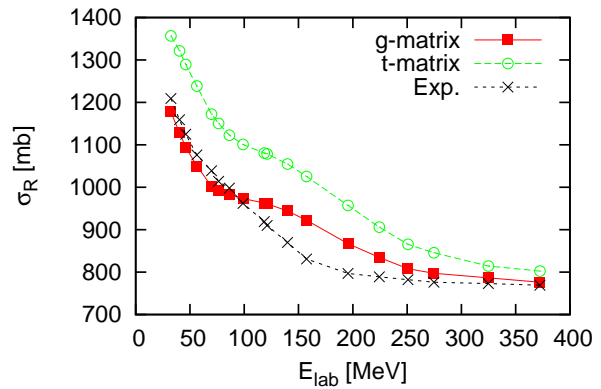}
\caption{
$E_{\rm lab}$ dependence of $\sigma_{\rm R}$ for $^{12}$C+$^{12}$C scattering. 
Closed squares stand for the $g$-matrix DFM results, 
while open circles correspond to the $t$-matrix DFM results densities. 
The experimental data (crosses) are taken from Ref.~\cite{Takechi:2009zz}. 
}
\label{fig:C+C}
\end{figure}

Figure~\ref{fig:C+C} shows $\sigma_{\rm R}$ as a function of  $E_{\rm lab}$ for  $^{12}$C+$^{12}$C 
scattering. Comparing our results with the data~\cite{Takechi:2009zz}, 
we confirm that the present DFM is reliable 
in $30  \lsim E_{\rm lab} \lsim 100 $~MeV and $250  \lsim E_{\rm lab} \lsim 400 $~MeV. 
 The $g$-matrix DFM results (closed squares) yield much better agreement with the experimental data 
 (crosses) than the $t$-matrix DFM results (open circles) do; note that 
 only the Kyushu and the Melbourne $g$-matrix approach the $t$-matrix, as $\rho$ becomes zero.  
 At $E_{\rm lab}=380$~MeV,  the $t$-matrix DFM result overestimates the data only by 4\%, 
 so that we may consider that the $t$-matrix DFM is accurate enough for $^{12}$C+$^{12}$C scattering 
 in $E_{\rm lab} \gsim 400$~MeV. 
 
\begin{figure}[H] 
\centering
\vspace{0cm}
\includegraphics[width=0.45\textwidth]{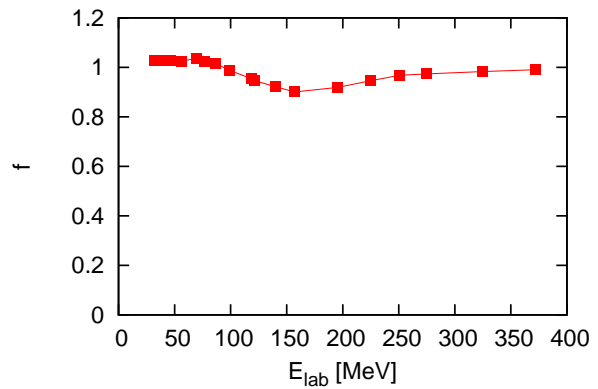}
\caption{
$E_{\rm lab}$ dependence of $f$ for $^{12}$C+$^{12}$C scattering. 
Closed squares stand for $f(E_{\rm lab}) \equiv \sigma_{\rm R}^{\rm exp}({\rm C+C})/\sigma_{\rm R}^{\rm th}({\rm C+C})$.  
}
\label{fig:factor}
\end{figure}

Figure~\ref{fig:factor} shows $E_{\rm lab}$ dependence of  
$f(E_{\rm lab}) \equiv \sigma_{\rm R}^{\rm exp}({\rm C+C})/\sigma_{\rm R}^{\rm th}({\rm C+C})$ for $^{12}$C+$^{12}$C scattering, where 
 $\sigma_{\rm R}^{\rm th}$ is the $g$-matrix DFM result. 
 The factor $|1-f|$ means an error of the present DFM, and becomes maximum 
around  $E_{\rm lab}=160$~MeV. Since the maximum error is still small, we guess that it comes from 
higher-order terms of chiral expansion for bare nucleon-nucleon force. Further explanation will be shown in 
Sec. \ref{Sec:summary} 

In order to minimize the error for other systems, we multiply  
``$\sigma_{\rm R}^{\rm th}({\rm other~system})$ 
calculated with   the $g$-matrix DFM'' by the factor $f(E_{\rm lab})$ and call the result ``the renormalized 
$g$-matrix DFM result'' from now on. 

Figure~\ref{fig:C+Be,Al} shows $E_{\rm lab}$ dependence of $\sigma_{\rm R}$ for $^{12}$C scattering on $^{9}$Be and $^{27}$Al targets. 
The renormalized $g$-matrix DFM results (closed squares) agree with the experimental data (crosses)~\cite{Takechi:2009zz}
within experimental error for  $E_{\rm lab} \gsim 75$~MeV. The renormalized $g$-matrix DFM results are reliable for 
 $E_{\rm lab} \gsim 75$~MeV.

\begin{figure}[H] 
\centering
\vspace{0cm}
\includegraphics[width=0.45\textwidth]{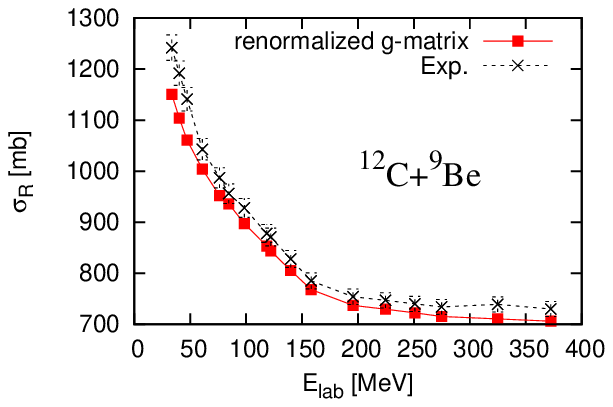}
\includegraphics[width=0.45\textwidth]{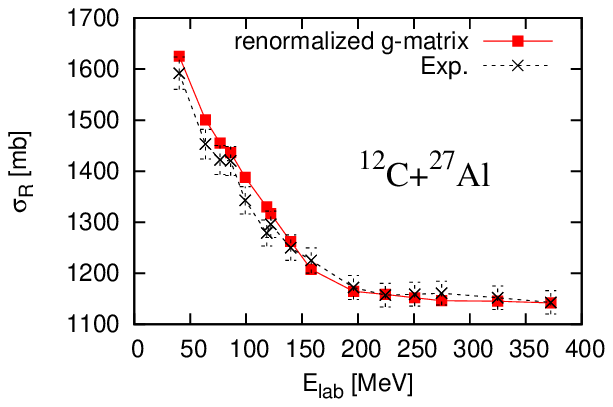}
\caption{
$E_{\rm lab}$ dependence of $\sigma_{\rm R}$ for $^{12}$C scattering on $^{7}$Be and$^{27}$Al targets. 
Closed squares show the renormalized $g$-matrix DFM results. 
The experimental data (crosses) are taken from Ref.~\cite{Takechi:2009zz} . 
}
\label{fig:C+Be,Al}
\end{figure}

\begin{figure}[H] 
\centering
\vspace{0cm}
\includegraphics[width=0.45\textwidth]{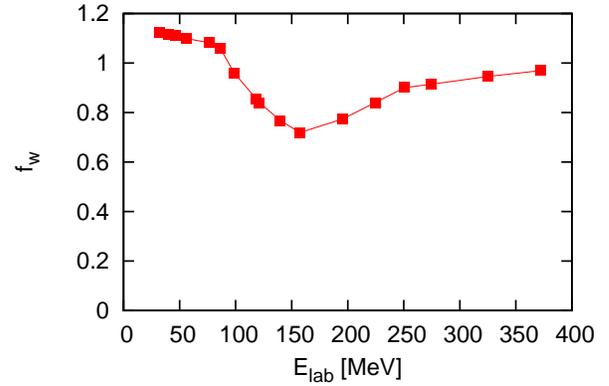}
\caption{
$E_{\rm lab}$ dependence of $f_{w}$ for $^{12}$C+$^{12}$C scattering. 
Closed squares stand for $f_w$.  
}
\label{fig:factor-w}
\end{figure}

\begin{figure}[H] 
\centering
\vspace{0cm}
\includegraphics[width=0.45\textwidth]{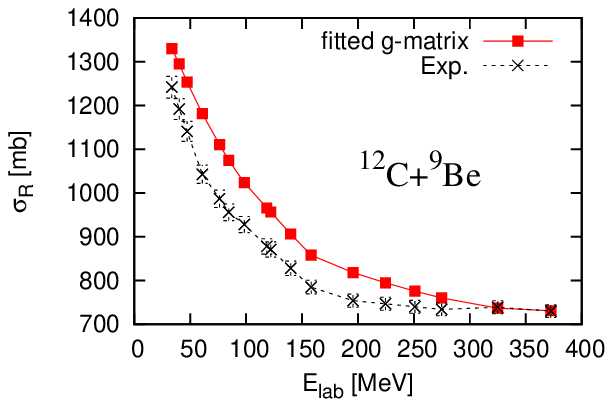}
\caption{
$E_{\rm lab}$ dependence of $\sigma_{\rm R}$ for $^{12}$C scattering on $^{9}$Be target. 
Closed squares show the fitted $g$-matrix DFM results. 
The experimental data (crosses) are taken from Ref.~\cite{Takechi:2009zz} . 
}
\label{fig:C+Be-fit}
\end{figure}

As an alternative prescription  to the  renormalized $g$-matrix DFM, 
we fit the imaginary part of the potential  ($g$ matrix) to the data on $\sigma_{\rm R}$ 
for $^{12}$C-$^{12}$C scattering. 
The fitting factor   $f_{w}$ is shown in Fig.~\ref{fig:factor-w}. The  $f_{w}$ tends to 1 as  
$E_{\rm lab}$ increases. Figure~\ref{fig:C+Be-fit} shows the results of DFM with the fitted $g$ matrix 
for  $^{12}$C scattering on a $^{9}$Be target. The fitted DFM well reproduces the data 
in  $E_{\rm lab} \gsim 300$~MeV. For the other $E_{\rm lab}$, the fitted DFM overestimates the data 
at most 13\%.

\section{Summary} 
\label{Sec:summary}

We predicted  the ground-state properties of Ca isotopes 
using GHFB and GHFB+AMP, 
and predicted the $\sigma_{\rm R}$ for scattering of Ca isotopes  on a $^{12}$C target 
at  $E_{\rm lab}=280$~MeV by using  the Kyushu $g$-matrix DFM~\cite{Toyokawa:2017pdd}. 
Details of the predictions are shown below. 

As an important property of Ca isotopes, we first determined the odd and even driplines 
by seeing $A$ dependence of $S_{1}$ and $S_{2}$, and found 
that  $^{64}$Ca is an even-dripline nucleus 
and $^{59}$Ca is an odd-dripline nucleus. 
As for $E_{\rm B}$ in addition to $S_{1}$, $S_{2}$, our results agree 
with the experimental data~\cite{HP:NuDat 2.7,Tarasov2018,Michimasa2018obr,Neufcourt:2019qvd} 
in $^{40-58}$Ca. Our results are thus accurate enough for the prediction on the odd and even driplines.  

As other grand-state properties of Ca isotopes, we considered  
$r_{\rm ch}$, $r_{\rm p}$, $r_{\rm n}$, $r_{\rm m}$, $r_{\rm skin}$, deformation for $^{40-60,62,64}$Ca. 
For $^{40-52}$Ca, the $r_{\rm ch}$ calculated with GHFB and GHFB+AMP are consistent with 
those~\cite{Angeli:2013epw}  deduced from the isotope shift method based on the electron scattering. 
As for $r_{\rm p}$, $r_{\rm n}$, $r_{\rm m}$,  $r_{\rm skin}$, the experimental data 
are available for  $^{48}$Ca~\cite{Birkhan:2016qkr}, and our results agree with  the data. 
The success mentioned above for the ground-state properties indicates that the densities calculated with 
GHFB and GHFB+AMP are reliable for Ca isotopes.

The  Kyushu $g$-matrix folding model is successful 
in reproducing the differential cross sections of $p$ scattering at 
$E_{\rm lab}=65$~MeV~\cite{Toyokawa:2014yma} and of $^4$He scattering at 
$E_{\rm lab}=30 \sim 200$~MeV~\cite{Toyokawa:2015zxa,Toyokawa:2017pdd}. 
However, it is not clear whether the Kyushu $g$-matrix DFM is reliable for $\sigma_{\rm R}$. 
We then applied the  Kyushu $g$-matrix DFM for measured $\sigma_{\rm R}$ on $^{12}$C scattering 
from $^{9}$Be, $^{12}$C, $^{27}$Al targets 
in $30  \lsim E_{\rm lab} \lsim 400 $~MeV, and 
confirmed that the Kyushu $g$-matrix DFM is reliable 
in $30  \lsim E_{\rm lab} \lsim 100 $~MeV and $250  \lsim E_{\rm lab} \lsim 400 $~MeV. 
We then predict $\sigma_{\rm R}$ for scattering of Ca isotopes  on a $^{12}$C target 
at  $E_{\rm lab}=280$~MeV, using the Kyushu $g$-matrix DFM. 
The reason for this prediction is that (1) 
the data on $\sigma_{\rm R}$ for $^{42-51}$Ca will be available soon and (2) 
the densities are determined accurately for Ca isotopes.  

The present DFM is not accurate enough for $^{12}$C-$^{12}$C scattering 
in  $110  \lsim E_{\rm lab} \lsim 240$~MeV. 
Whenever we use the chiral interaction, $E_{\rm lab}$ should be smaller than $\L=550$~MeV. 
In general, the chiral $g$-matrix DFM becomes less accurate as $E_{\rm lab}$ increases. 
The small error in  $110  \lsim E_{\rm lab} \lsim 240$~MeV seems to come from terms higher than 
the present order. The reason why the present DFM  is good for higher $E_{\rm lab}$ is 
that the present $g$-matrix approaches the  $t$-matrix as $E_{\rm lab}$ increases. 
In order to improve the present DFM in $110  \lsim E_{\rm lab} \lsim 240$~MeV, 
we have proposed two prescriptions. 
The renormalized DFM proposed is good for  $30  \lsim E_{\rm lab} \lsim 400 $~MeV.
The values of the present $g$-matrix is published in Ref.~\cite{Toyokawa:2017pdd}
and the homepage http://www.nt.phys.kyushu-u.ac.jp/english/gmatrix.html. 
For $E_{\rm lab} \gsim 400 $~MeV, we recommend the $t$-matrix DFM.

\section*{Acknowledgements}
We thank Dr. M. Toyokawa and Prof. Y. Iseri heartily. 
The authors express our gratitude to  Dr. Y. R. Shimizu for his 
useful information.


\end{document}